\documentclass{emulateapj}

\usepackage{epsfig}

\slugcomment{In Preparation for the Astrophysical Journal}

\shorttitle{Water Content of Low Mass Star Planets}
\shortauthors{Fred J. Ciesla}

\begin{document}


\title{Volatile Delivery to Planets from Water-rich Planetesimals around Low Mass Stars}

\author{Fred J. Ciesla}
\affil{Department of the Geophysical Sciences, The University of Chicago, 5734 South Ellis Avenue, Chicago, IL 60637}

\author{Gijs D. Mulders, Ilaria Pascucci}
\affil{Lunar and Planetary Laboratory, University of Arizona, Tucson, AZ 85721}

\author{and D\'aniel Apai}
\affil{Lunar and Planetary Laboratory and Steward Observatory, University of Arizona, Tucson, AZ 85721}

\newpage

\begin{abstract}

Most models of volatile delivery to accreting terrestrial planets assume that the carriers for water are similar in water content to the carbonaceous chondrites in our Solar System.  Here we suggest that the water content of primitive bodies in many planetary systems may actually be much higher, as carbonaceous chondrites have lost some of their original water due to heating from short-lived radioisotopes that drove parent body alteration.  Using N-body simulations, we explore how planetary accretion would be different if bodies beyond the water line contained a water mass fraction consistent with chemical equilibrium calculations, and more similar to comets, as opposed to the more traditional water-depleted values.  We apply this model to consider planet formation around stars of different masses and identify trends in the properties of Habitable Zone planets and planetary system architecture which could be tested by ongoing exoplanet census data collection.  Comparison of such data with the model predicted trends will serve to evaluate how well the N-body simulations and the initial conditions used in studies of planetary accretion can be used to understand this stage of planet formation.

\end{abstract}

\keywords{}

\newpage

\section{Introduction}

Low mass stars are attractive targets for the search for habitable worlds due to their great abundance in the galaxy \citep[nearly 75\% of stars are M dwarfs,][]{scalo07}, and the fact that planetary detection signals from radial velocity, transits, direct imaging and microlensing would be greater for these stars than those with higher mass and greater size.  As such, recent efforts have been dedicated to predicting the characteristics of planets that may form around such stars in order to evaluate how common habitable planets are in the galaxy \citep[e.g.][]{raymond07,lissauer07,ogihara09,montgomery09}.

One of the primary objectives in these investigations has been to evaluate the likelihood that planets around these stars may have accreted water during their formation and thus be potential abodes for life.
\citet{raymond07} and \citet{lissauer07} examined this possibility in detail, considering whether the same processes which we believe were responsible for the delivery of water to the young Earth could have also operated around these lower mass stars.  That is, Earth is  expected to have accreted largely from native, dry planetesimals and embryos as temperatures were too high at 1 AU in the solar nebula for water to have condensed as a solid.  Earth's oceans would thus have come from water-bearing solids which formed outside the water line--the distance from the young Sun where water ice was able to condense and be incorporated into solid planetesimals.  The accretion of such materials from more distant regions is a natural consequence of the dynamics of planetary accretion in our Solar System \citep{morbidelli00,obrien06,raymond09}, and is consistent with the similar D/H ratio observed in Earth's oceans and chondritic meteorites whose parent bodies now populate this region of the Solar System \citep{alexander12,marty12}.

\citet{raymond07} first explored how much water-bearing material was accreted by habitable zone (HZ) planets that formed around stars of different masses.  Scaling the location of the water line with the main sequence luminosity of the star, \citet{raymond07} found that while the frequent gravitational encounters between bodies beyond the water line  around solar-mass stars would lead to some of these materials being scattered inward to be accreted by the inner planets, such encounters were less frequent and less intense during planet formation around low-mass stars.    As a result, little to no water-bearing bodies would be scattered inwards, leaving the planets in the habitable zones around these stars to accrete predominately native, dry materials, thus leaving them volatile poor.

\citet{lissauer07} pointed out an additional difficulty for the accretion of volatiles by HZ planets around low-mass stars.  Specifically, the slow contraction of low mass stars means that they would be more luminous during the epoch of planet formation than during their main sequence lifetime.  This increased luminosity is important as irradiative heating by the central star plays an important role in setting the location of the water line in a protoplanetary disk, particularly when accretional heating through the midplane of the disk is negligible.  Thus, while \citet{raymond07} scaled the location of the water line with the main sequence luminosity of the star, \citet{lissauer07} argued that the water line would be located even further from low-mass stars than this scaling would suggest.  Thus water-bearing solids would have to be scattered inward from greater distances in order to be accreted by HZ planets around very low mass stars.  Given the findings of \citet{raymond07}, such scattering would still have been minimal given the low surface density of solids expected in the disk, again leaving HZ planets water-poor.

An important assumption in these models is that the water-bearing planetesimals and embryos beyond the water line in all of these systems were similar in composition to carbonaceous chondrites, the water-bearing primitive bodies which are thought to be the source of Earth's water \citep{morbidelli00, obrien06, raymond09, alexander12}.  In our Solar System, these bodies formed through the accretion of a mix of rocky minerals and water ice beyond the water line.  Once they accreted, the decay of short-lived radionuclides (SLRs), most notably $^{26}$Al, raised the internal temperatures of the bodies, melting the water ice and allowing it to react with the rock to form hydrated minerals \citep{grimm89,cohencoker00,castillorogez10}.  This process was also capable of mobilizing or driving off water, meaning that the final water content seen in rocky meteorites is less than the original accreted mix.  In fact, under equilibrium conditions in the solar nebula, the water-to-rock mass ratio beyond the water line should be nearly 1-to-1 \citep{lodders03}; however carbonaceous chondrites contain only up to 5-10\% water by mass, suggesting possible loss of this water after accretion.  Thus in other planetary systems where $^{26}$Al may have been less abundant or absent,  bodies beyond the water line could have been much more water-rich and the total surface density of solids greater than previously considered in planetary accretion studies.

Here we explore how a greater water content of distant bodies would affect planetary accretion around stars of different masses.  Specifically, we build on the studies of \citet{raymond07} and \citet{lissauer07} and focus on  how the absence of live SLRs during planet formation, which could be the norm for many planetary systems, affects the accretion of planets in such systems and the final water content of potentially habitable planets around low mass stars.  In the next section we discuss in greater detail the water content of planetesimals in protoplanetary disks and the location where water is expected to be incorporated into solid bodies.  We follow that by presenting a series of simulations which compare how planetary accretion occurs under the standard assumption of using carbonaceous chondrites as analogs for the bodies originating beyond the water line and with how it occurs when those bodies are composed of equal parts water and rock.  We end by discussing our results and highlighting areas in need of future work.

\section{Water distribution in planetesimals}

Water condenses in a protoplanetary disk where its partial pressure exceeds the saturation pressure, $P_{H_{2}{O}}^{disk} > P_{H_{2}O}^{sat}$.  While the exact temperature where this occurs depends on the details of the physical structure of the disk and relative abundances of elements, the  expected temperature ranges from 140-180 K for typical disk conditions.  Here, we will adopt a value of $T_{cond}$=140 K to guide our calculations, recognizing that the results of our study are not very sensitive to the exact choice of condensation temperature.  Further, choosing a value at the low end of the expected range means that the results presented here are conservative, with higher temperatures meaning the water line would be located closer to the star, making it easier for water-containing bodies to be delivered to HZ planets.

The temperature structure of our protoplanetary disk, the solar nebula, throughout its evolution is uncertain, and it likely evolved with time as the physical properties of the Sun and nebula itself evolved.  Nonetheless, the water line of the solar nebula has typically been thought to have left its mark at $\sim$2.5 AU, as C-type asteroids, which are thought to be the parent bodies of carbonaceous, water-bearing meteorites, are the dominant type of meteorite found outside this distance, while dry, S-type asteroids, are the main bodies found interior to this distance \citep{gradietedesco82,morbidelli00,raymond05,raymond07,raymond09}.   However, it is worth noting that recent observations indicate that this zoning is only true for the largest objects, and that asteroid compositions exhibit greater variations when smaller objects are considered \citep{demeo14}.  This has been interpreted as evidence that 
 the positions of the asteroids today are not the same as their formation location, and instead the asteroids were scattered into their current positions as the planets were assembled, something that has been suggested by a variety of dynamical models \citep[e.g.][]{bottke06, walsh12}.  If so, then the distribution of asteroid compositions today cannot be used to provide strong constraints on the thermal structure of the solar nebula.

In practice, the temperature within a protoplanetary disk, and thus the location of the water line, is determined by balancing the heat inputs from irradiation from the central star, the ambient radiation field, and viscous dissipation within the disk with the energy lost by radiation at the disk surface.  A key issue in determining the thermal evolution of the disk is identifying the magnitude and location where viscous dissipation occurs. If dissipation occurs throughout the entire thickness of the disk, temperatures everywhere will be increased because of the added energy.  However, it is unclear whether dissipation would occur throughout, or if it would be limited to particular regions of the disk.  For example, if the MRI is responsible for driving disk accretion, viscous heating may be limited to 
 regions of high ionization fractions like the extreme inner edge of the disk, the disk surface layers, or the outer disk \citep[e.g.][]{gammie96}.  In such a case, irradiation from the central star is likely the dominant source of heating in the regions where planets formed, and thus would set the location of the water line \citep{lecar06,martinlivio12}.  
 
 When irradiative heating is dominant, the midplane temperature at some distance, $r$, from the central star can be calculated as \citep{chiang97}:
\begin{equation}
T \left( r \right) = \left( \frac{\alpha}{4} \right)^{\frac{1}{4}} \left( \frac{R_{*}}{r} \right)^{\frac{1}{2}} T_{*}
\end{equation}
where $T_{*}$ is the effective temperature of the star, $R_{*}$ is the radius of the star,  and $\alpha$ is the grazing angle of the radiation on the disk surface.  The grazing angle will likely depend on location in the disk, as it is determined by the height of the dust layers in the disk which will evolve with time as dust grows and settles to the midplane.  Despite these details, $\alpha$ is expected to have a value of $\sim$0.05 in the regions of the protoplanetary disk where the water line is expected \citep{chiang97}, with slight differences being unimportant as temperature is proportional to $\alpha^{\frac{1}{4}}$--an angle that is 50\% larger would lead to a temperature change of just $\sim$15\%.  Again, if viscous dissipation within the disk were taking place throughout the disk, then temperatures would be greater than this, and thus the temperatures given by this equation should be considered a minimum value.  

Figure 1 illustrates how the location of the water line evolves in a disk as a result of the pre-main sequence contraction of stars of different masses.  Here the stellar tracks from \citet{siess00}, which calculate the evolution of the radius and temperature of stars of different masses over time, were used along with Equation 1 to determine the region in the disk where $T$=140 K.  Note that other stellar models would likely give slightly different numbers, though the trend of cooling with time are robust.  The change in water line location is most rapid early in stellar evolution as the star contracts at the greatest rate.  

As planetary accretion simulations determine the provenance of the different building blocks for a planet in order to determine its final composition, the location of the water line marks a transition from dry planetesimals and embryos to those that are water-bearing.  Previous models for the accretion of the terrestrial planets have largely used carbonaceous chondrites as analogs for the bodies originating just beyond the water line and thus have assigned a water content of $\sim$5\% to those bodies coming from that region \citep[e.g.][]{morbidelli00,obrien06,raymond05,raymond07,raymond09}.  Indeed,
carbonaceous chondrites contain 5-10\% water by mass, though this water is not stored as pure water or ice, but instead within hydrated silicates.  These minerals formed on the meteorite parent bodies, as the decay of $^{26}$Al heated the parent body, melting any water ice which was present (melting temperature $\sim$ 273 K, thus $>$100 K higher than the nebular condensation temperature).  The liquid water then reacted with the anhydrous rock, forming minerals such as serpentine, brucite, and cronstedtite \citep[e.g.][]{grimm89,cohencoker00,castillorogez10,hewins14,zolotov14}.

Thus it is important to remember that the $\sim$5\% mass fraction of carbonaceous chondrites we see today is the product of the geophysical and geochemical evolution of the original parent body which formed from a mix of rock and ice over 4.5 billion years ago.  It is likely that the original parent bodies contained more water than seen in the carbonaceous chondrites today.  The same radiogenic heat which melted the water ice and allowed hydrated minerals to form also could have mobilized the water, allowing it to migrate to the surface of the asteroid where it was lost to space or frozen out in the surface layers where it would later sublimate or be eroded by impacts\citep{grimm89,cohencoker00,young03,castillorogez10}.   In fact, cronstedtite, which is found in many hydrated meteorites \citep[e.g.][]{hewins14} is thought to require very high water-to-rock ratios (water-to-rock mass ratio $>$1-10, or even higher) to form \citep{zolotov14}.  Further, based on the mineralogy of CO carbonaceous chondrites, \citet{howard14lpsc} has estimated that these meteorites formed in bodies that had a water mass fraction of $\sim$50\%, with water being lost through oxidation reactions and degassing that occurred during the early thermal evolution of the parent asteroid.  Such mass fractions are not unheard of in asteroidal bodies, as the largest body in the asteroid belt, Ceres, is believed to contain at least 20-30\% water by mass \citep{mccord05,castillorogez10}.    Much of this water is expected to be found as the outer mantle of the asteroid, which formed as a result of differentiation of the asteroid \citep{castillorogez10}.  The outer layers would have been subjected to loss of water by sublimation and impact erosion, suggesting that the water seen in Ceres today is only a lower limit on the amount which was present when the asteroid first formed.  
 
These high water mass fractions are consistent with
chemical equilibrium models for the composition of solids in the solar nebula which predict that solids beyond the water line should be $\sim$50\% water by mass \citep{lodders03}.  Indeed, such ratios are seen in other bodies in the outer solar system, with comets and icy satellites having water mass fractions of $\sim$40-50\% \citep[e.g.][and references within]{delsemme92,morbidelli00,schubert10,ahearn11}.    However, even higher values are possible as this estimate ignores any dynamic effects which could occur within a protoplanetary disk.  Diffusion of water vapor in the protoplanetary disk and inward drift of icy solids would combine to concentrate water ice just beyond the water line \citep{stevenson88,cyr98,stepinski97,hueso03,cuzzi04,cieslacuzzi06}.  The enhancement of water in this region could lead to even greater water mass fractions than expected under what is predicted under chemical equilibrium.

Thus, here we  explore how water-rich planetesimals from immediately beyond the water line would impact the accretion and chemical properties of the terrestrial planets that form around a star.  Water loss from planetesimals in our solar system was likely due to radiogenic heating, notably the decay of $^{26}$Al, whose abundance was likely variable across different planetary systems.  \citet{vasileiadis13} demonstrated that the $^{26}$Al abundance in giant molecular cloud cores will increase in time as early formed stars introduce freshly synthesized short-lived radionuclides to the star-forming region.  Thus early-formed planetary systems would be lacking the radionuclides to drive thermal evolution on planetesimals, while those that formed later could be similar to our own Solar System, allowing similar thermal evolution of small bodies to take place \citep{jura13,young14}.  However, the efficiency with  which radionuclides would be injected and mixed into a forming planetary system is uncertain, and could be rare or require special circumstances \citep[e.g.][]{boss10,oullette10,williams07,adams14}.  As a result, it is necessary to consider how the absence of such radiogenic heating could preserve water-rich planetary building blocks beyond the snow line, making them more comet-like in their water-to-rock ratios than carbonaceous chondrite-like, and the consequences for planetary accretion.

\section{Planetary Accretion Simulations}

In order to explore the effects that water-rich planetesimals would have on planetary accretion, we performed 40 simulations using the {\tt MERCURY} N-body code \citep{chambers99}.  As the primary goal here is to understand how the presence of water-rich planetesimals would affect the volatile inventory of planets around different stars, we
follow a similar approach as \citet{raymond07} to facilitate comparisons between our findings and previous results.  That is, we consider stars ranging in mass from $M_{*}$=0.2-1.0 $M_{\odot}$ and assume that the surface density of rocky planetary building blocks is described by:
\begin{equation}
\Sigma \left( r \right) = \Sigma_{1} \left( \frac{r}{\mathrm{1~AU}} \right)^{-1} \left(\frac{M_{*}}{M_{\odot}}\right)
\end{equation}
where $\Sigma_{1}$ is the surface density at 1 AU, and $M_{*}$ is the mass of the star of interest.   Following \citet{raymond07}, we adopt a value of $\Sigma_{1}$=6 g cm$^{-2}$ for all cases considered.  Note that we are assuming here that the metallicity and relative elemental abundances of the planetary systems are identical to those of our Solar System--the impact of variations in elemental abundances \citep[e.g.][]{johnson12} on the properties of planets is left to future studies.  For each system, we assume the solids in the disk are distributed from an inner radius, $R_{in}$, to an outer radius $R_{out}$.  While the choice of such parameters are largely arbitrary in these models, we scale $R_{out}$ to correspond to the location where $\Sigma \left( R_{out} \right)$=0.75 g cm$^{-2}$.  $R_{in}$ was chosen to be roughly half the distance inward of the habitable zone.  This differs from the approach of \citet{raymond07} who scaled the inner and outer boundaries of the solids disk with the main sequence luminosity of the star, though the inner boundaries used here and in that study are nearly identical, while the outer edges of the disks are generally at greater distances in our simulations.

  Where \citet{raymond07} considered all of the solid mass in the disk to be contained within planetary embryos, here we distribute 50\% of this mass into embryos and 50\% into planetesimals to represent the initial stages of oligarchic growth \citep{kokuboida98, obrien06,raymond09}.  As in previous studies, embryos interact gravitationally and collisionally with both other embryos and planetesimals, while planetesimals only interact with the embryos.  We assume that the initial mass of the embryo is given by $M_{emb}$=$ \frac{1}{20} \left( \frac{M_{*}}{M_{\odot}} \right) M_{\oplus}$; that is, for a solar mass star, the embryos have a mass roughly half of that of Mars, with this mass decreasing proportionally with mass of the star.     In all cases, the mass of planetesimals in our simulation is equal to $\frac{1}{20}$ that of the embryos.  Previous models have varied in how they have treated the initial masses of the embryos: some use variable embryo masses spaced by a multiple of their mutual Hill radii \citep[e.g.][]{raymond07,raymond09} while others have used embryos of constant mass spaced in a way to mirror the surface density of the solids of the disk \citep[e.g.][]{obrien06,lissauer07}.  While details about individual runs are known to vary with the exact initial conditions of an N-body run, the trends in regards to the final outcomes (planet number, masses, volatile content) that emerge when examining these different studies are independent of the choice of how embryo properties are initialized.  Thus we expect the results described here to be independent of the choice of embryo mass distribution used: constant embryo mass or variable embryo mass.

We consider two  scenarios for each star.  First, we follow previous studies in assuming that the building blocks in our models are made predominately out of rock, and that beyond the water line the bodies are composed of 5\% water.  We label these simulations as CC (for `carbonaceous chondrite').  We also consider situations where the water content of the materials originating beyond the water line is 50\%, and label these simulations as IC (for `icy').  In these later cases, the surface density of materials beyond the water line, ($r>r_{WL}$) is doubled to reflect the increase in the mass of available solids there.

In all cases, the water line is assumed to occur at $T$=140 K, with its position is calculated using Equation (1) and adopting the stellar parameters for 1 Myr as given by \citet{siess00}.  The location of the water line, along with other model parameters, in each case is given in Table 1.  Note that the relative locations of the water line in the different mass stars remain nearly constant after 1 Myr, except for the lowest mass star. The water line sets the boundary between the 'dry' and 'wet' planetesimals and embryos in each simulation, as these bodies are expected to reflect the composition of their local environment when they form within the gaseous disk given that the planetesimals grow from the dust present in the disk and embryos grow from the planetesimals.   Note that the water line is expected to be much closer to the central star in the higher mass cases than was used in \citet{raymond07}.  This is  because the temperature profile of the disk is determined by the reprocessing of radiation at the disk surface downward.   For the lower mass stars, however, the water line is located much further out than in \citet{raymond07} as the luminosities of the stars during their early evolution are much greater than their main sequence luminosities, even offsetting the indirect radiation geometry caused by the disk, as discussed by \citet{lissauer07}.

The choice of setting the location of the snow line at 1 Myr is somewhat arbitrary.  As the stars continue to contract with time, the location of the water line will migrate inwards under the assumptions used here, as illustrated in Figure 1.   In our own Solar System, there is evidence that planetesimal \citep{kruijer14} and embryo \citep{dauphas11} formation took place $<$1 Myr into its evolution, while some meteorites provide evidence of parent body formation 2-5 Myr later \citep{krot05, kleine09}.  Formation of Jupiter via core accretion likely requires relatively rapid planetesimal formation given that envelope accretion takes millions of years, possibly longer than the median lifetime of protoplanetary disks \citep{hubickyj05}.  Thus planetesimal and embryo formation likely spanned millions of years within the solar nebula, meaning the boundary between dry and wet planetesimals was likely not static, but evolved with time, and this extended period may have impacted the chemistry of the forming planets \citep{moriarty14}.  Here we follow previous studies by using a singular location for the water line to represent an average location.  As discussed here and in previous studies, these results should be used to be a general guide on how planetary accretion would occur under the conditions described here as every system likely varied in terms of the stages of evolution that preceded the stage studied here.

\section{Results}

Figure 2 shows the dynamical evolution of the embryos and planetesimals for one of our 0.4$M_{\odot}$ CC cases, while Figure 3 shows a simulation for one of the IC cases around a star of the same mass.  Accretion in each of these simulations proceeds as has been documented in previous studies: the orbits of planetesimals and embryos are excited by gravitational interactions with one another leading objects to be put on more eccentric, and thus crossing, orbits.  The crossing of the orbits leads to the formation of larger bodies as a result of the collisions that occur.  In general, planetesimals achieve more excited orbits as seen by the higher eccentricities of these smaller bodies in Figures 2 and 3.  This dynamical friction has been seen in previous studies \citep[e.g.][]{obrien06} and means that these small bodies are much more likely to be accreted by planets which formed at different semi-major axes than the planetesimals.  Thus these planetesimals are likely to carry materials across greater radial distances than the embryos, and deliver materials formed at more distal locations than the planets.

This last point is important in considering the differences in the planetary systems which result in the final panels of Figures 2 and 3.  In the CC case, the inner most planets have negligible water content as they accreted largely from native materials, that is materials that originated in the same region where those planets reside at the end of the simulation.  Even with the inclusion of planetesimals here, this result is in agreement with the findings of \citet{raymond07}, who also found that planets which formed close to the star around 0.4$M_{\odot}$ mass stars would largely be composed of dry materials from the inner regions of the solid disk.  This contrasts with the results of the IC case, where the inner most planets have water mass fractions of 1.2-3.3\%.  As discussed further below, the ability for inner planets to accrete water-bearing planetesimals is due to the greater mass of material present beyond the water line in the IC cases compared to the CC cases.  In the case of the two inner most planets, their water content was reached by accreting one single planetesimal each from beyond the water line in addition to the native materials it accreted during their growth.  The third planet in the system actually accreted 6 planetesimals from beyond the snow line, leading to a greater water abundance.  Unlike the first two planets, the third planet was also formed by the merger of two embryos, those being the third and fourth embryos when ranked in terms of initial semi-major axis at the beginning of the simulation.  These embryos were both dry, having originated inside the water line, meaning that while the third planet accreted 6$\times$ the mass of water-rich planetesimals, the fraction of water in the planet was only 3$\times$ that of the inner most planets. 

These trends continue when we consider the full suite of simulations around stars of different masses.  Figure 4 shows the final planetary systems that were formed in each of the cases considered here, with stars of 0.2, 0.4, 0.6, 0.8, and 1.0 $M_{\odot}$.   While some planetesimals remain unaccreted at the end of each simulation, and thus could potentially be accreted at later times (much like the planets in our Solar System today continue to accrete extraterrestrial materials), these planetesimals have largely diffused to larger semi-major axes and thus are less likely to have a significant impact on the planets which formed close to the star, in the habitable zone region.  In each case shown, the four CC simulations are shown in the bottom of the panels, while the IC cases are shown above them.  The Solar System is shown at the top of each plot for reference, with the water mass fractions for the planets set at 10$^{-5}$ for Mercury, 3.7$\times$10$^{-5}$ for Venus \citep{donahue97}, $\sim$0.001 for Earth \citep[a lower limit as more water may be contained in the Earth's mantle][]{morbidelli00}, and 3.5$\times$10$^{-4}$ for Mars \citep{lunine03}.  Again, we stress that the Solar System is provided for reference--our goal here is not to reproduce the properties of our own Solar System, but rather to explore the diversity of planetary systems that may develop and how this is impacted by the enhancement of water ice beyond the water line in different planetary systems.

Among the key differences that are seen in these simulations are that the water content of the innermost planets around each star are greater in the IC cases when compared to the traditional CC cases.  These wetter planets are due to the increase in mass beyond the water line in the CC cases.  Firstly, the fact that the water content of each planetesimal in the IC case is 10$\times$ greater than the traditional water content of planetesimals considered in the CC case means the accretion of just one single body from beyond the water line would deliver significantly more water than considered in previous work.  Further, the extra mass beyond the water line (in the IC case there is 2$\times$more mass than in the CC case) results in more frequent and stronger gravitational interactions among solids in this region, leading to more significant scattering of planetesimals.  This allows such bodies to undergo much more radial mixing than in the CC cases.  We note that as in previous studies, we are assuming perfect mergers in these simulations and not allowing for the potential water loss that could occur during accretion.

It is worth noting that while the CC cases considered here result in drier HZ planets than the IC cases, these planets tend to contain more volatiles than found in the study by \citet{raymond07}.  In particular, here the  HZ planets that form around the 0.6$M_{\odot}$ stars contain volatiles in the CC cases, while HZ planets which formed around the 0.8$M_{\odot}$ stars were fairly dry in \citet{raymond07}.  This difference is due to the initial conditions of these simulations including 50\% of the solid mass in planetesimals, which were neglected in the previous studies.  This equal-mass distribution of embryos and planetesimals are consistent with more recent simulations of planetary accretion \citep[e.g.][]{raymond09,chambers13}.  As the water-rich bodies in the previous studies tended to be much more massive than considered here, they did not experience significant radial migration due to gravitational scattering events.  Here, however, the low mass planetesimals are easily excited to high eccentricity orbits, as shown in Figures 2 and 3, allowing water-bearing bodies to be accreted by forming planets over a wide range of semi-major axes.  Thus planetesimals, or smaller, low mass solids from beyond the water-line are potentially the critical carriers of water to HZ planets around M stars.

A second important trend that can be seen in these simulations is that the number of planets that form in these simulations increases around the smaller mass stars.  This trend was also seen in the results of \citet{raymond07}, and arises due to the increased available mass at the beginning of the simulations and the greater level of gravitational interactions around the higher mass stars.  That is, as described above, planetary accretion arises due to the orbital excitation of their building blocks which leads to collisions.  This excitation occurs much more readily in the higher mass disks due to the greater masses of bodies involved.  This also leads to the greater number of planets around low mass stars, as the lower masses in these disks do not lead to significant orbital excitation of the embryos and planetesimals.  This trend is also seen in the Kepler data as there have been greater numbers of planets found around individual low-mass stars than high mass stars \citep{howard12,mulders14}.  In addition, the orbital spacing between the planets in the low mass star cases is less due to the lower levels of gravitational interactions.   This trend is also seen in comparing our CC cases to the IC cases--there are a greater number of planets formed in the CC cases due to the lower masses of the disks in these simulations, while the increased mass beyond the water line in the IC cases allows for more orbital excitation of the planetary building blocks.

\section{Discussion and Conclusions}

Here we have shown that the increase in solid surface density in a forming protoplanetary system due to the condensation of water ice can have an important effect on the properties of the planets which form around a young star.  The increase in solid surface density leads to more massive planets and planets with greater water contents than would be found in systems where water-bearing planetesimals are treated as carbonaceous chondrite analogs.  These effects are not limited to the regions immediately outside the water line--gravitational scattering means that the supply of planetesimals to regions inside the water line, including the eventual habitable zone, will be increased.  This allows for more massive and volatile-rich planets to form closer to the star than previously realized.  

The findings here are generally in agreement with previous studies of the volatile content of habitable zone planets around low mass stars \citep{raymond07,lissauer07}.  Indeed, the cases where we followed traditional assumptions by not increasing the surface density of solids at the water line, we found dry planets in the habitable zones for stars with masses $<$0.6$M_{\odot}$ as in previous work.  However, in those cases where the solid surface density was increased at the water line (the IC cases), we find that dry habitable zone planets may be limited to only very low mass disks or stars $\sim$0.2$M_{\odot}$ or less.

The results of these types of models depend sensitively on the choice of initial conditions used in these simulations, such as the water line location or surface density of planetesimals.  In order to avoid exploring a large parameter space, we limited ourselves to only one surface density profile for the building blocks of the planets and linearly scaled the surface density of the planetary building blocks with the central mass of the star.  This linear scaling is consistent with the protoplanetary disk mass scaling relationship reported by  \citet{andrews13}, though significant variations among disks demonstrate such scaling should be used as a guide and not a definite rule.  This issue was also discussed by \citet{raymond07}, and like those authors, we suggest that these results be taken as a statistical guide for inferring the properties of planets around stars of different masses.

While we did not model a range of disk masses around a single stellar mass, we can use the results presented here to draw some inferences about how planetary properties may vary for disks with masses different from those considered here.  One trend to note is that the ranges in the properties of the habitable zone planets also scale with mass of the central star.  Figure 5 shows how the masses and water mass fractions of the planets in the habitable zones of the respective stars vary in the runs presented here.  For the lowest mass stars considered, 0.2$M_{\odot}$ and 0.4$M_{\odot}$, we find that the masses of the planets in the habitable zone are tightly clustered at values that are just slightly in excess of the mass of the embryos used in those simulations, with variations of only $\sim$10-20\%.  That is, these planets generally represent single embryos which accreted a number of planetesimals and avoided impacts with other embryos.  This was true in both the CC cases and IC cases, as the planets that formed in the habitable zones around these stars were sufficiently far from the location of the water line that the increase in surface density in the further reaches of the disk did not have a dramatic effect on their growth.  The more massive stars, however, have habitable zone planets whose masses range by factors of 2-20.  Likewise, the water mass fraction of habitable zone planets cluster at minimal values for the low mass stars (all have WMF=10$^{-5}$ in the 0.2$M_{\odot}$ case), but the range increases with increasing mass.  The range in WMF is actually greater in the CC cases than the IC cases for stars $>$0.6$M_{\odot}$, likely because the fraction of water-bearing building blocks is significantly greater in the IC cases, meaning individual accretion events can change the water inventory of a planet more significantly than in the CC cases. 

The increase in the ranges in planetary mass with increasing stellar mass seen in the simulations is a product of the chaotic nature of planetary accretion caused by the gravitational encounters of a number of randomly placed embryos.  In the low stellar mass cases, gravitational encounters among planetary embryos are less frequent and less intense due to the fact that mass is less available and less concentrated in these low mass disks.  Thus planetesimal-embryo or embryo-embryo collisions and scattering will not occur as frequently as in disks with higher masses, allowing for more orderly growth.  Thus, the range in properties of the planets seen in these models is likely a product of the disk properties and less so the properties of the star.  This suggests that planets that form from more massive disks are likely to display a wider range of characteristics than those that form from lower mass disks.  Further, if this extrapolation is correct, terrestrial planets which form from low mass disks are likely to serve as fossil records of the the properties of the protoplanetary disk--the low levels of gravitational scattering among planetary building blocks mean that planets which form in such disks would grow from planetesimals and embryos that formed locally.

An important caveat to this work is that the effects of more massive planets, like gas or ice giants and super-Earths, have been neglected.  In our own Solar System, different orbital properties of Jupiter and Saturn would have led to significantly different accretion histories for the inner planets, differences which go beyond those attributed to the stochastic nature of planetary accretion \citep{obrien06,raymond09,fischer14,quintana14}.  In addition, migration of gas giants could play a critical role in determining how terrestrial planets accrete in a young planetary system  \citep{raymond06,walsh11}.  Thus such planets, if present, could lead to different accretion outcomes and different relationships between the habitable zone planets and the stellar mass than found here.  Again, we neglected such effects in order to focus on the consequences of the increased surface density at the water line.  However, gas giants are not present in all systems; in fact, their occurrence rate may be as low as 10\%, meaning most systems are likely to be born without any such planets present \citep{kasper07,howard12,mortier12,zechmeister13,biller13,dressing13}.

Super-Earths and ice giants are present around stars at higher frequencies than gaseous giants, though still are not present in all systems \citep{howard10,dressing13,fressin13}.  The formation of these objects remains uncertain.  It is possible that such objects form through accretion of solids, as outlined here, coupled with radial migration through the gaseous disk \citep{raymond08,raymond14}.  Indeed, depending on the rate of migration, the inward movement of accreting bodies could play a important role in setting the structure of the planetary system and chemical composition of the planets that form \citep[e.g.][]{ogihara09}.  Indeed, the effects of planetary migration on planetary accretion is likely variable due to uncertainties on drift rates and will likely vary depending on turbulence levels and the mass of the disk at the time accretion begins \citep{chambers09}.

Others have suggested that super-Earths formed \emph{in situ} through gravitational accumulation of planetesimals and embryos as described here \citep{chianglaughlin13}.  Indeed, \citet{kennedy06} explored the formation of such planets using analytic models for accretion in the context of the framework described here; that is, such planets were thought to possibly form outside the water line of a protoplanetary disk where the surface density of solids was increased by a factor of 2 or 3 due to the condensation of water ice as considered here.  We have shown that one would expect more massive planets to form when such a surface density increase is accounted for, though to get very massive planets would likely require much more massive disks than assumed here.  Again, this should be the focus of future work.

Thus the detailed results here are most applicable to those planetary systems which form from relatively low mass disks and where massive planets are absent. Despite the additional effects that could be considered in the future, this work has shown that the increase in surface density and high water-content of bodies expected beyond the water line of a protoplanetary disk can have an important effect on the accretion of terrestrial planets.  Specifically, the increase in mass beyond the water line will lead to more gravitational scattering than in those cases where bodies are treated as carbonaceous chondrite analogs, leading to larger and more volatile-rich planets than seen in previous studies.  The water-rich planetary building blocks considered here are likely the norm in planetary systems which contained little to no short-lived radionuclides.  Even in our own solar system, planetary building blocks which originated beyond the water line were likely more water-rich than the carbonaceous chondrites we see today as aqueous alteration, fluid flow, and water loss would have taken millions to tens of millions of years to occur \citep{grimm89,cohencoker00,young03,castillorogez10}.  Thus during the period of planetary assembly, these bodies likely looked different than previously assumed, allowing for water-rich bodies to be more common than had been realized.  The effects of having more water-rich parent bodies for the carbonaceous chondrites in our own Solar System need to be investigated.

In considering the suite of results presented here, along with those described in \citet{raymond07}, the general trends that emerge are interesting to consider as we collect more data on the properties of extrasolar planets and their planetary systems.  As discussed above, these simulations predict two major trends to develop: planetary systems will contain greater numbers of lower mass planets in small orbital spacing around low mass stars while higher mass stars will be home to fewer, more massive planets.  In addition, within the simulations for a single stellar mass, the predicted planetary properties follow general trends showing how water mass fraction would vary with semi-major axis, as shown in Figure 6.  Thus, even though every planetary system that forms in these studies is different, due to the different assumed positions of the initial building blocks, the resulting suite of planets define a radial trend.  Such predictions provide an opportunity to test N-body accretion studies, or the choices of initial conditions used by such studies, against actual data.  That is, as shown in previous work, the outcome of individual N-body simulations are highly dependent on the initial conditions, meaning that individual runs are unlikely to reproduce all of the properties of any given planetary system.  However, taking the collection of simulations together and comparing them to the suite of exoplanetary systems, specifically the H$_{2}$O/rock ratio as determined by mass-radius relationships \citep[e.g.][]{seager07,rogers10,swift12}, could allow us to test our ideas of this stage of planet formation in greater detail.   Should the trends predicted here be consistent with the planetary systems that are detected, then this would give support that our choices for the initial conditions in these N-body simulations are accurately describing those from which planetary systems develop.  Should we see trends that are inconsistent with these predictions, that would suggest that the way in which we initialize our N-body simulations may not accurately describe the conditions under which planetary accretion begins, requiring other possibilities to be considered in detail.  In fact, trends such as these may allow us to infer the location of the water line in protoplanetary disks as shown in Figure 6--the water line was located at 0.95 AU in these simulations and one readily sees that this point marks a difference between planets that have nearly uniform water mass fractions (outside the water line, and thus the plateau in this plot) and those that have decreasing water contents as one gets closer to the star (inside the water line).  Alternatively, disagreement in data and the model predictions could indicate that the processes ignored in N-body simulations, such as inelastic collisions and perfect mergers or disk-driven migration, are actually important and must be considered in accretion studies.  Thus the growing collection of constraints on the numbers, masses, and orbital properties of extrasolar planets now allows us to explicitly test our models for planet formation. Such efforts will help us to better understand the origin of our own Solar System.

\emph{Acknowledgements.}  The authors are grateful to the very useful comments and suggestions made by Sean Raymond which led to a more complete and clearer manuscript.  This work was supported by the NASA Exobiology program through grant NNX12AD59G. This material is partly based upon work supported by the National Aeronautics and Space Administration under Agreement No. NNX15AD94G, Earths in Other Solar Systems, issued through the Science Mission Directorate interdivisional initiative Nexus for Exoplanet System Science.

\bibliographystyle{apj}


\newpage
\begin{table}
\begin{center}
Table 1: Simulation Parameters\\
\quad

\begin{tabular}{c c c c c c c c c c}
\hline
& Initial & 
Water & 
& 
 & 
 & &  & & Habitable\\
$M_{*}$  &  Range & Line & $M_{emb}$ & $M_{plan}$ & $N_{emb}^{CC}$ & $N_{plan}^{CC}$  & $N_{emb}^{IC}$ & $N_{plan}^{IC}$  & Zone \\
($M_{\odot}$) & (AU) & (AU) & ($M_{\oplus}$) & ($M_{\oplus}$) & & & & & (AU) \\
\hline

1.0 & 0.5-4.0 & 1.3 & 0.05 & 2.5$\times$10$^{-3}$ & 50 & 974 & 88 & 1875 &  0.8-1.5\\
0.8 & 0.2-3.2 & 1.1 & 0.04 & 2$\times$10$^{-3}$ & 43 & 827 & 72 & 1528 & 0.39-0.74\\
0.6 & 0.1-2.4 & 0.9 &  0.03 & 1.5$\times$10$^{-3}$ & 33 & 622 & 54 & 1122 & 0.20-0.37\\
0.4 & 0.05-1.6 & 0.8 &  0.02 & 10$^{-3}$ & 23 & 420 & 34 & 672 & 0.10-0.19\\
0.2 & 0.03-0.8 & 0.5 & 0.01 & 5$\times$10$^{-4}$ & 11 & 196 & 16 & 291 & 0.05-0.10\\
\hline

\end{tabular}
\end{center}
$M_{smb}$: Embryo mass; $M_{plan}$: Planetesimal mass; $N_{emb}$: Number of embryos, $N_{plan}$: Number of planetesimals.  Habitable Zone estimates taken from \citet{raymond07}
\end{table}

\newpage
\begin{figure}
\begin{center}
\includegraphics[width=5in,angle=90]{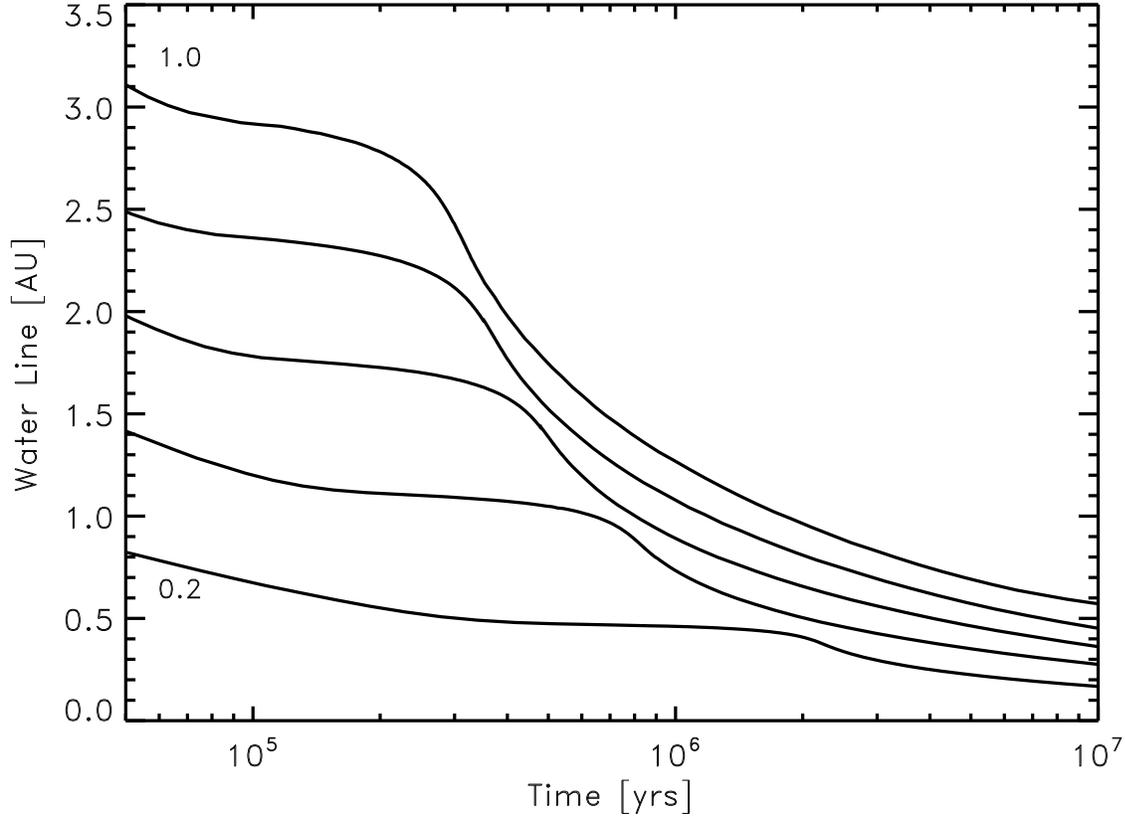}
\end{center}
\caption{Location of the water line, defined by $T$=140 K, as a function of time for 0.2, 0.4, 0.6, 0.8, and 1.0 $M_{\odot}$ stars based on the models of \citet{siess00}. }
\end{figure}

\newpage
\begin{figure}
\begin{center}
\includegraphics[width=5in,angle=90]{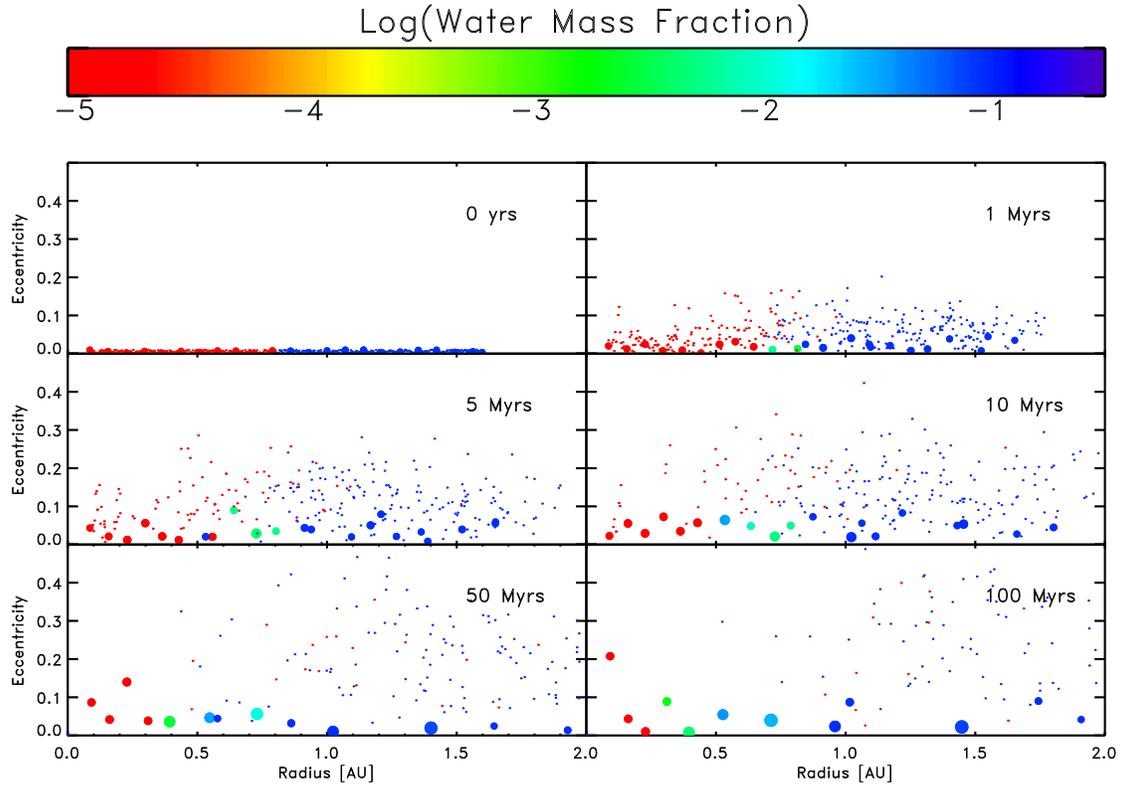}
\end{center}
\caption{Series of snapshots showing the accretion history of one of our 0.4$M_{\odot}$ CC cases.  Symbol size scales with the mass of the bodies shown, while colors indicate the corresponding water mass fraction.}
\end{figure}

\newpage
\begin{figure}
\begin{center}
\includegraphics[width=5in,angle=90]{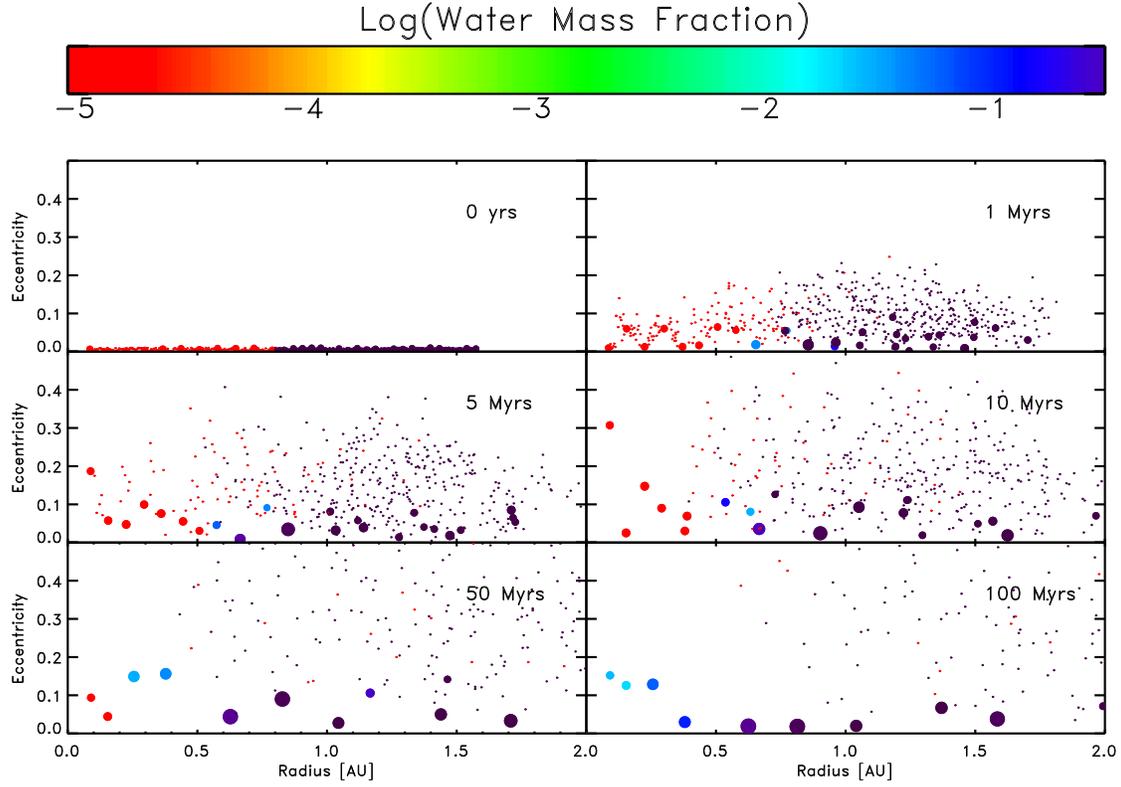}
\end{center}
\caption{Series of snapshots showing the accretion history of one of our 0.4$M_{\odot}$ IC cases.  Symbols and colors are as shown in Figure 2.  An important difference to note here is that the water mass fraction extends to higher values here than in the CC cases.  Also note that the surface density of bodies beyond the water line in this simulation is greater than the CC case.  This leads to more dynamical excitement of bodies in the outer disk and, thus, more radial mixing of planetary building blocks. }
\end{figure}

\newpage
\begin{figure}
\includegraphics[width=2.6in,angle=90]{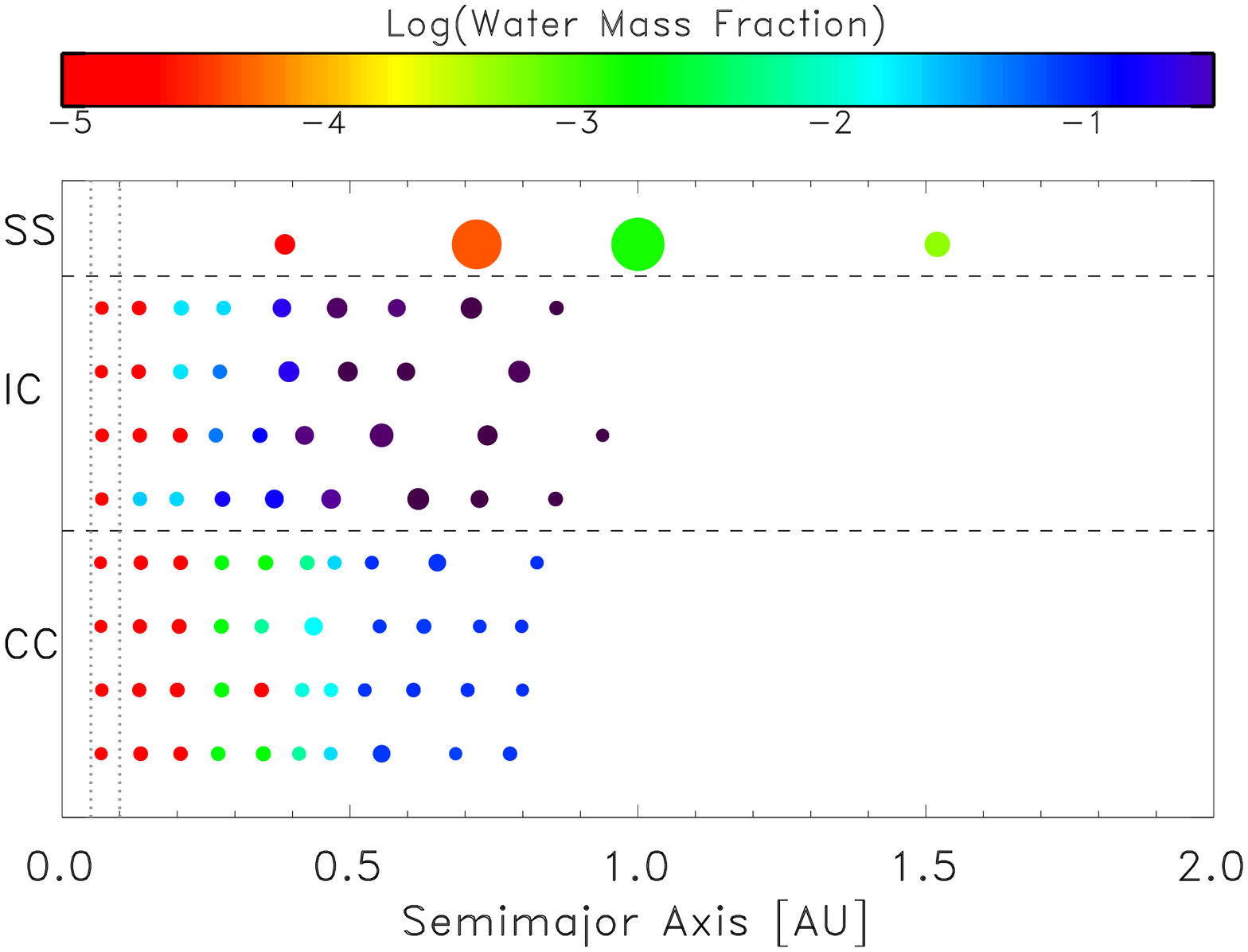}
\includegraphics[width=2.6in,angle=90]{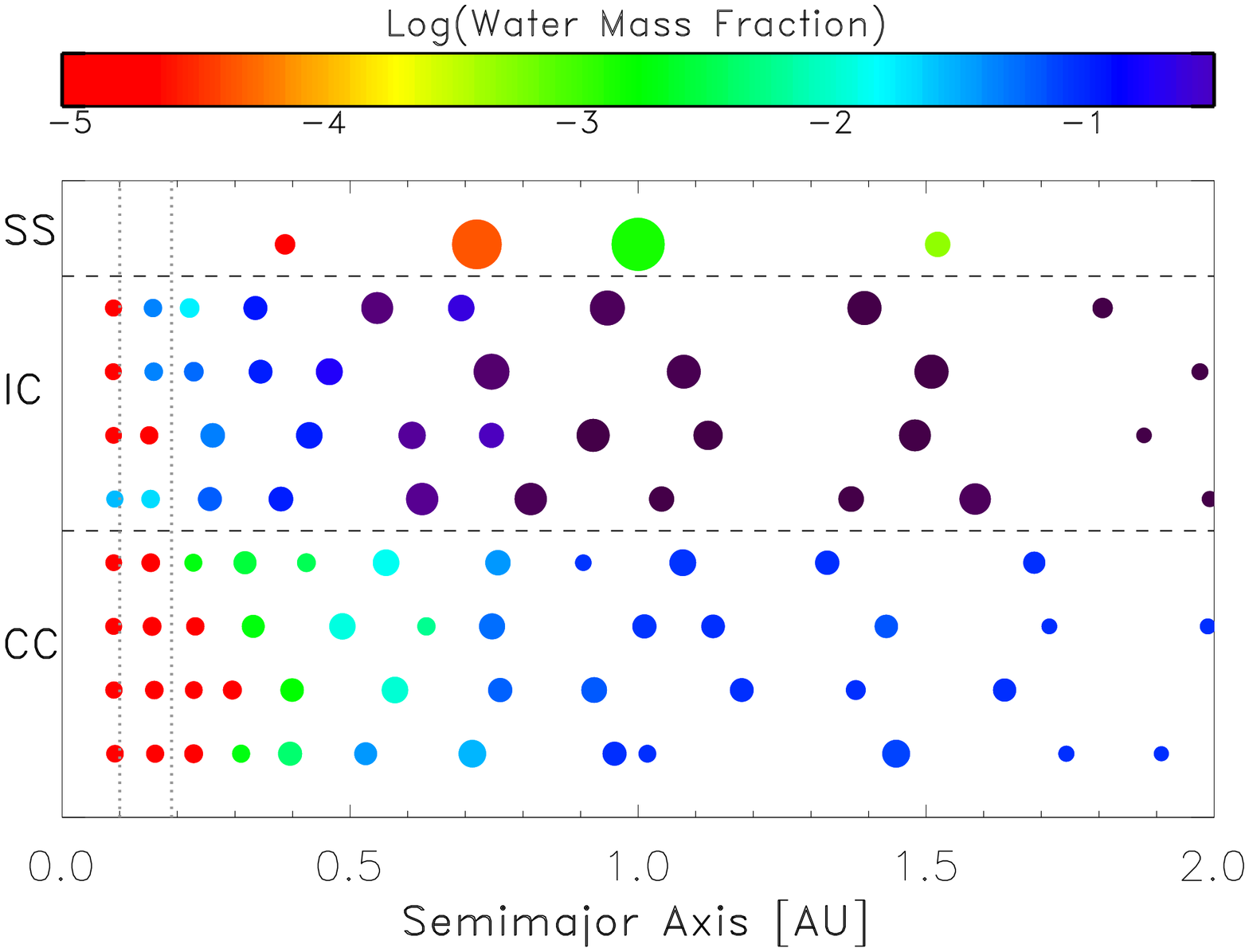}
\includegraphics[width=2.6in,angle=90]{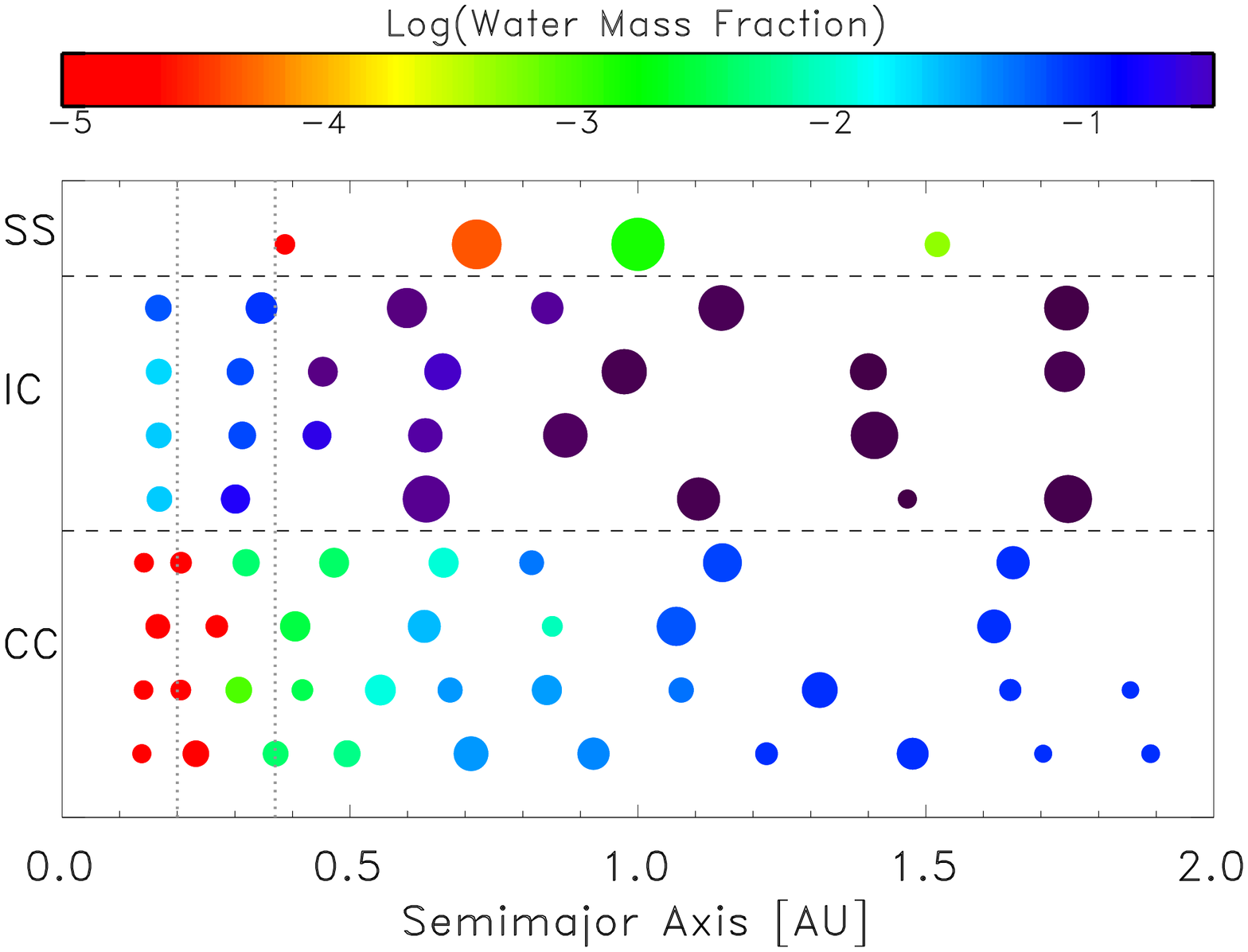}
\includegraphics[width=2.6in,angle=90]{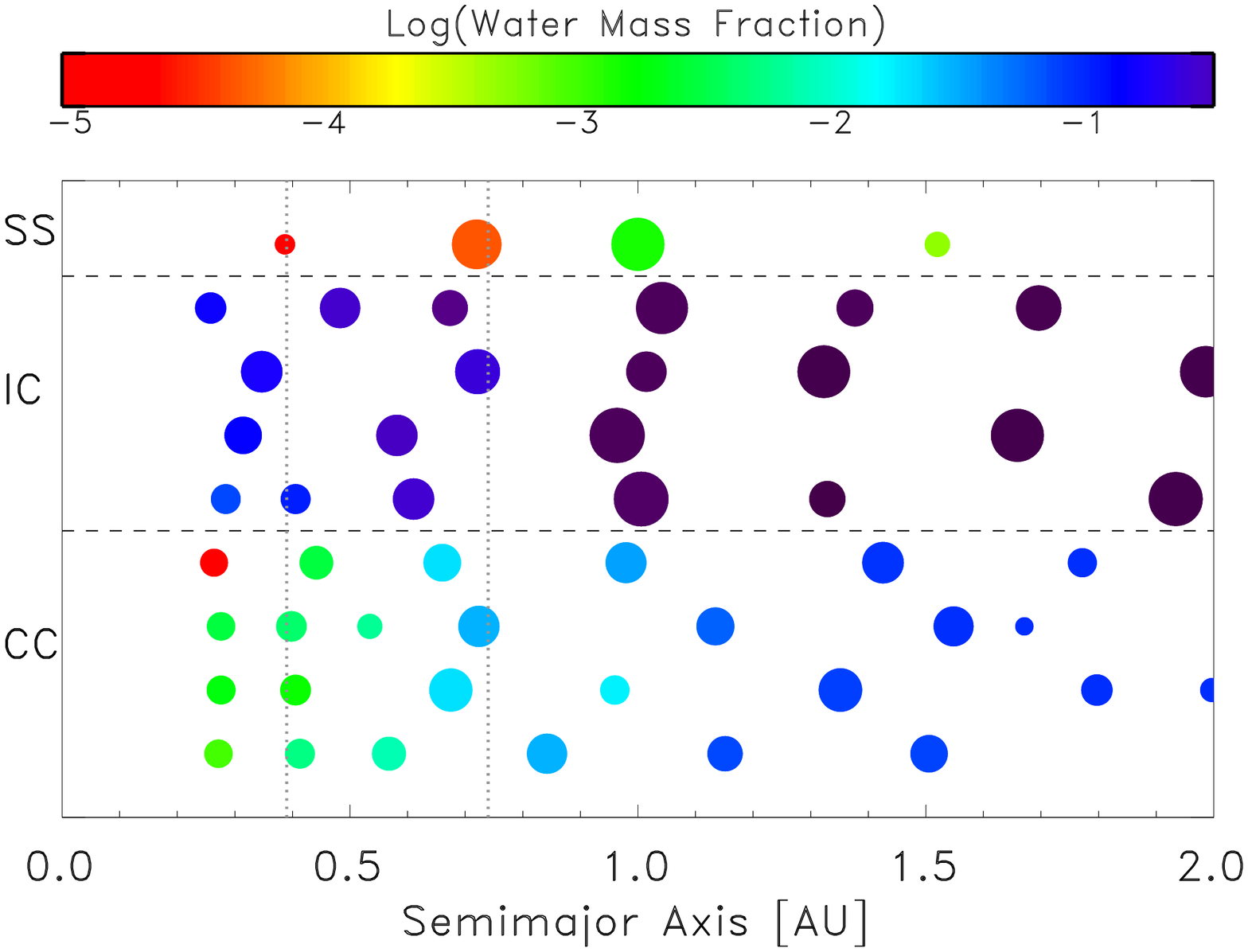}
\includegraphics[width=2.6in,angle=90]{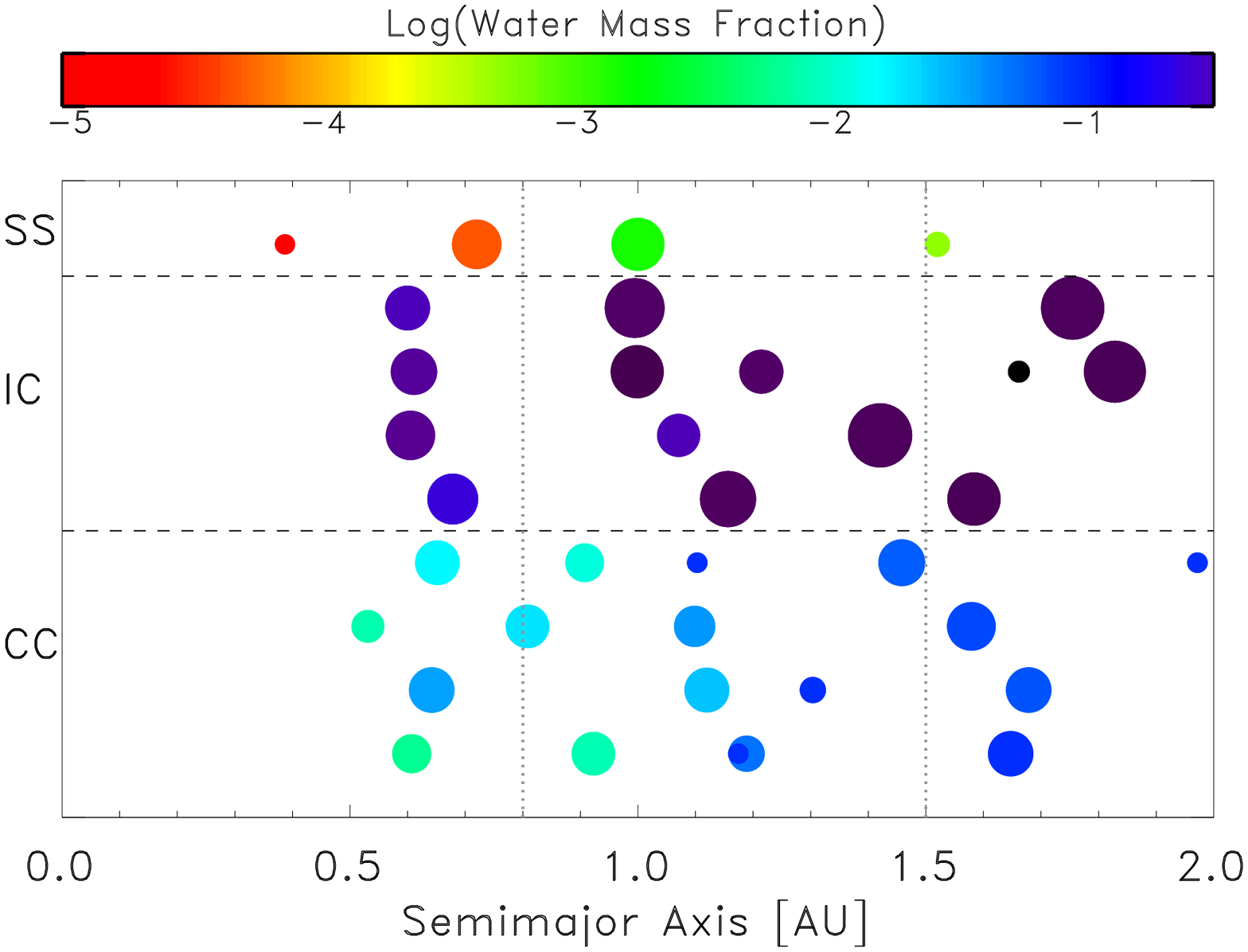}
\caption{Final architectures of all simulations done in this study, in order of increasing stellar mass.  Vertical lines indicate the location of the Habitable Zone as given in Table 1.  As in Figures 2 and 3, symbol size scales with planetary mass to the $\frac{1}{3}$ power, while symbol color indicates the water mass fraction of the planet.  Traditional, CC cases are shown in the bottom of each simulation, while the more water-rich, IC cases are shown above them.  The Solar System planets are shown for reference in each panel.}
\end{figure}

\newpage
\begin{figure}
\begin{center}
\includegraphics[width=3in,angle=90]{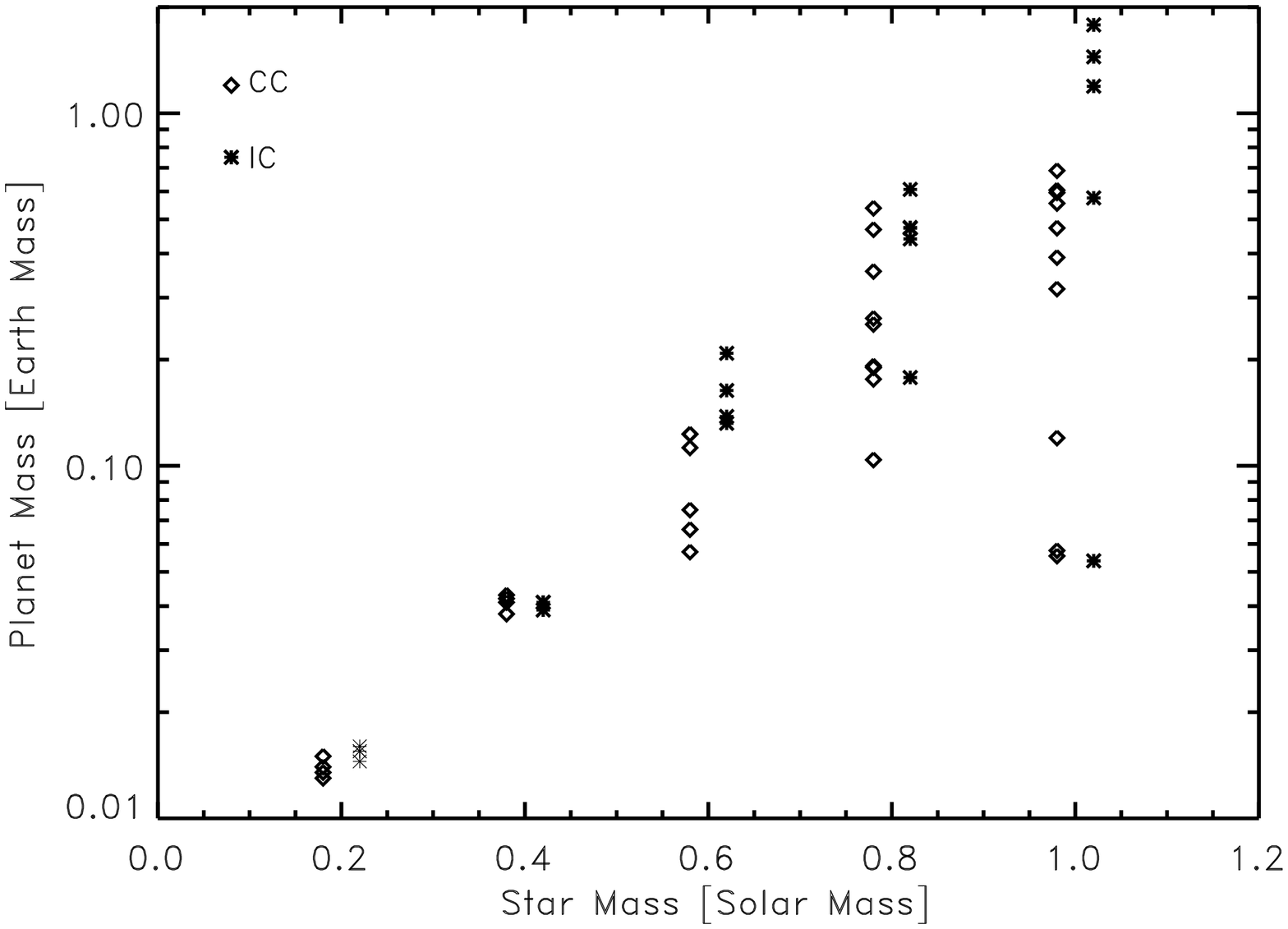}
\includegraphics[width=3in,angle=90]{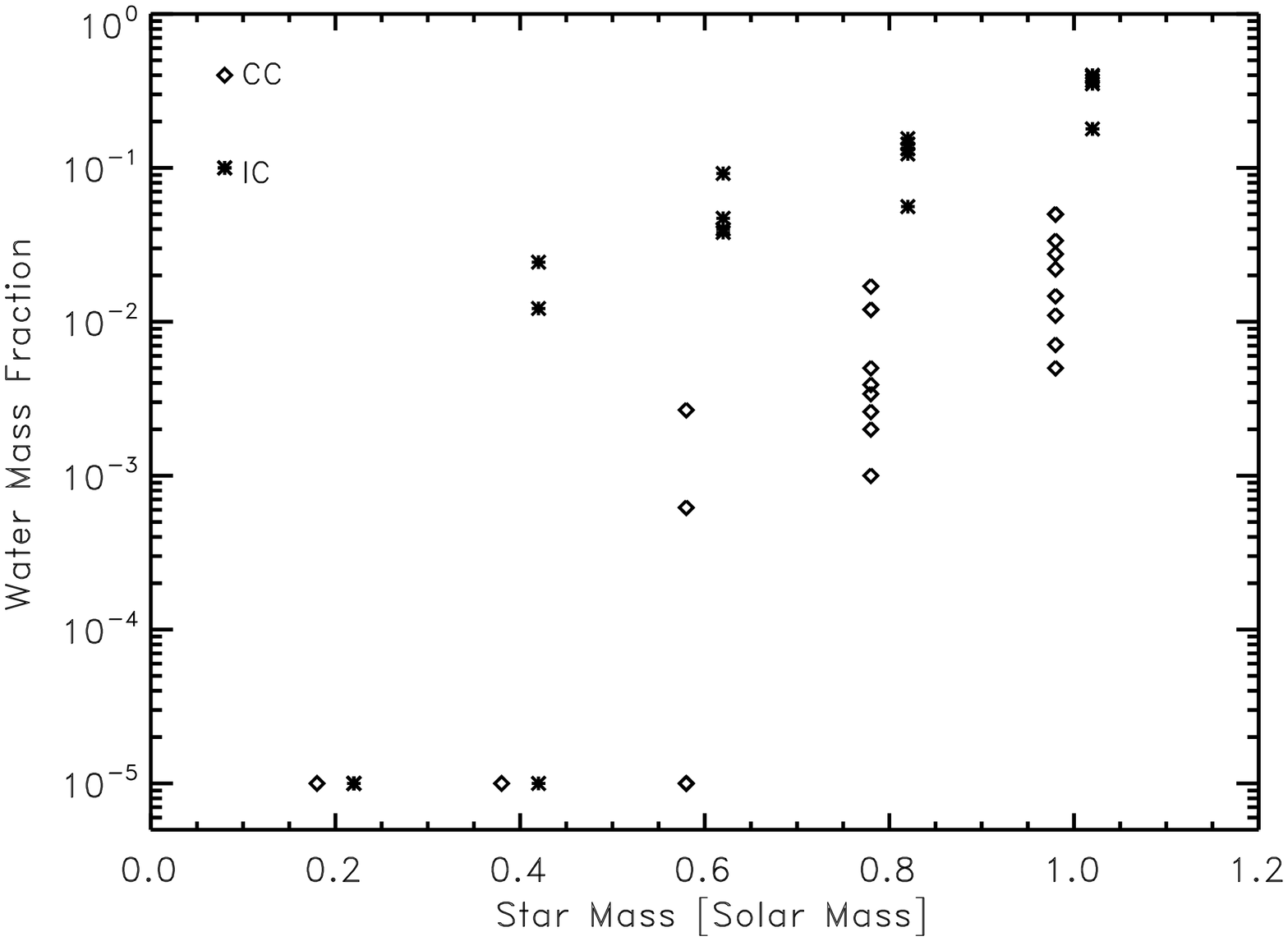}
\end{center}
\caption{Properties of only those planets which reside in the Habitable Zone for each star at the end of the simulations as a function of stellar mass.  Top panel shows the masses of the Habitable Zone planets, and how such planets exhibit greater ranges in values for higher mass stars.  This is due to the more extensive gravitational interactions that occur in the higher mass disks which accompany such stars, and thus the more chaotic nature of this evolution.  Growth around lower mass stars is much more orderly.  The lower panel shows the water mass fractions for the Habitable Zone planets in this simulation, showing that the low mass stars, due to the low masses of the disks around them, do not see significant inward scattering of water-rich bodies during planetary accretion, while inward scattering is occurs more readily in the more massive disks (either the IC cases or higher mass stars).}
\end{figure}

\newpage
\begin{figure}
\begin{center}
\includegraphics[width=5in,angle=90]{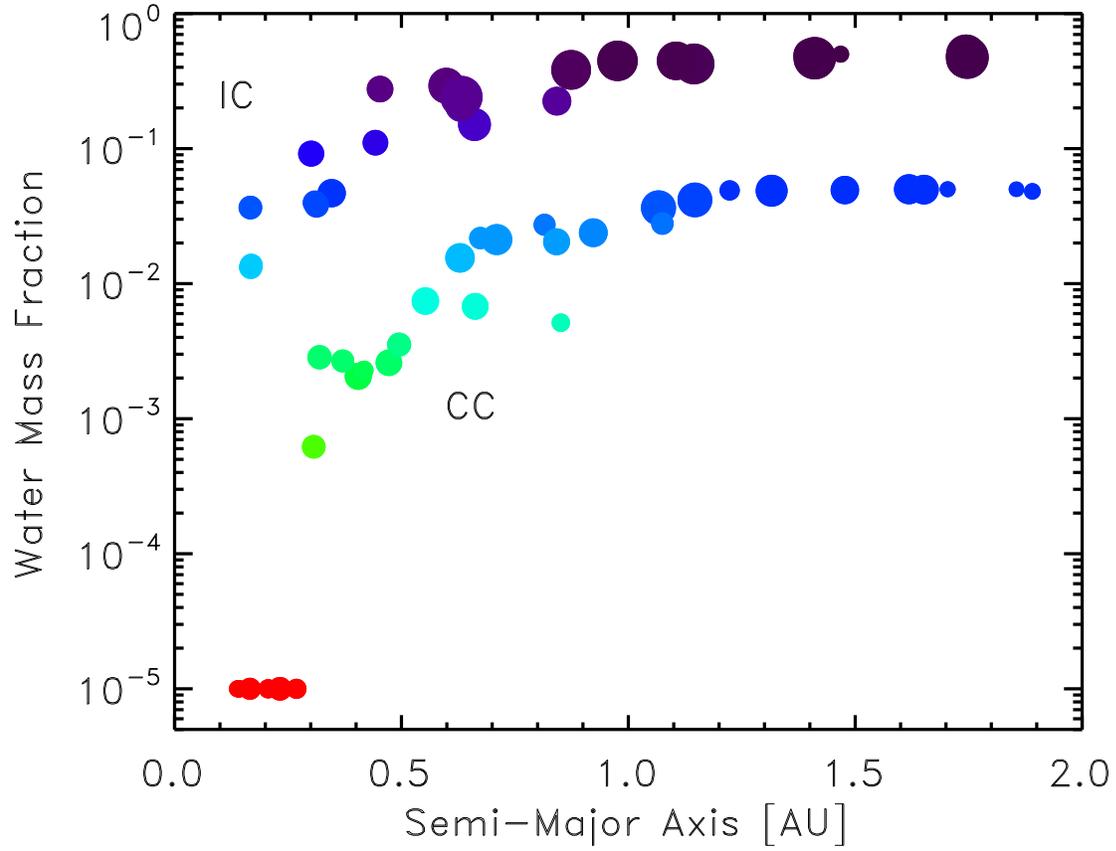}
\end{center}
\caption{Cumulative distribution of planetary properties for the 0.6$M_{\odot}$ runs performed in this study, showing how planetary properties vary as a function of semi-major axis.  While every simulation produced different planetary systems, a trend that develops in these simulations is the radial gradient in the water mass fraction of the planets that arise from radial diffusion of water-bearing planetesimals due to gravitational interactions beyond the water line.  Looking for trends like these in the collection of exoplanetary systems would serve as important tests for our models and ideas on the final stages of planetary accretion.}
\end{figure}

\end{document}